\documentclass[aps,prl,twocolumn,showpacs,amsmath,amssymb,superscriptaddress]{revtex4}

\usepackage{graphicx}  
\usepackage{dcolumn}   
\usepackage{bm}        
\usepackage{amssymb}   
\usepackage[usenames]{color}
\usepackage[colorlinks,urlcolor=blue,citecolor=blue,linkcolor=blue]{hyperref}

\begin{document}
\newcommand{\bra}[1]{\mbox{\ensuremath{\langle #1 \vert}}}
\newcommand{\ket}[1]{\mbox{\ensuremath{\vert #1 \rangle}}}
\newcommand{\mb}[1]{\mathbf{#1}}
\newcommand{\phipp}{\big|\phi_{\mb{p}}^{(+)}\big>}
\newcommand{\phipav}{\big|\phi_{\mb{p}}^{\p{av}}\big>}
\newcommand{\pp}[1]{\big|\psi_{p}(#1)\big>}
\newcommand{\drdy}[1]{\sqrt{-R'(#1)}}
\newcommand{\Rb}{$^{87}$Rb}
\newcommand{\kf}{$^{40}$K}
\newcommand{\na}{${^{23}}$Na}
\newcommand{\muK}{\:\mu\textrm{K}}
\newcommand{\p}[1]{\textrm{#1}}
\newcommand\T{\rule{0pt}{2.6ex}}
\newcommand\B{\rule[-1.2ex]{0pt}{0pt}}
\newcommand{\reffig}[1]{\mbox{Fig.~\ref{#1}}}
\newcommand{\refeq}[1]{\mbox{Eq.~(\ref{#1})}}
\hyphenation{Fesh-bach}
\newcommand{\previous}[1]{}
\newcommand{\note}[1]{\textcolor{red}{[\textrm{#1}]}}

\preprint{}

\title{Radio-frequency dressing of multiple Feshbach resonances}

\author{A. M.~Kaufman}
\affiliation{Department of Physics, Amherst College, Amherst, MA 01002-5000 USA}

\author{R. P.~Anderson}
\affiliation{Department of Physics, Amherst College, Amherst, MA 01002-5000 USA}
\affiliation{ARC Centre of Excellence for Quantum-Atom Optics and Centre for Atom Optics and Ultrafast Spectroscopy, Swinburne University of Technology, Hawthorn, Victoria 3122, Australia}

\author{Thomas M. Hanna}
\affiliation{Joint Quantum Institute,
  NIST and University of Maryland,
  100 Bureau Drive, Stop 8423,
  Gaithersburg, MD 20899-8423, USA}

\author{E. Tiesinga}
\affiliation{Joint Quantum Institute,
  NIST and University of Maryland,
  100 Bureau Drive, Stop 8423,
  Gaithersburg, MD 20899-8423, USA}

  \author{P. S. Julienne}
    \affiliation{Joint Quantum Institute,
  NIST and University of Maryland,
  100 Bureau Drive, Stop 8423,
  Gaithersburg, MD 20899-8423, USA}

  \author{D.~S.~Hall}
\affiliation{Department of Physics, Amherst College, Amherst, MA 01002-5000 USA}

\begin{abstract}
We demonstrate and theoretically analyze the dressing of several proximate Feshbach resonances in $^{87}$Rb using radio-frequency radiation (rf). We present accurate measurements and characterizations of the resonances, and the dramatic changes in scattering properties that can arise through the rf dressing. Our scattering theory analysis yields quantitative agreement with the experimental data. We also present a simple interpretation of our results in terms of rf-coupled bound states interacting with the collision threshold.
\end{abstract}

\date{\today}


\maketitle

The precise manipulation of ultracold atomic collisions underlies many recent advances in diverse areas of physics, including metrology~\cite{cronin09}, many-body physics~\cite{bloch08}, and quantum information theory~\cite{sorensen01}. Such control can be achieved with magnetically or optically tunable Feshbach resonances~\cite{chin_review}. Independent modification of the interactions between different species in a multicomponent gas is an important building block for realizing more complicated processes such as Efimov physics~\cite{dincao09} and color superfluidity~\cite{rapp07}, but has so far remained elusive due to the absence of concurrent mechanisms of control.

The direct modification of atomic scattering lengths with radio-frequency radiation (rf) has generally been considered impractical due to small Franck-Condon overlaps between bound and scattering states. Recent work has nevertheless shown that rf is useful for dissociating~\cite{regal03} and associating~\cite{thompson05, ospelkaus06, beaufils08} Feshbach molecules, as well as for driving transitions between bound states~\cite{lang08b}. These experiments imply that the simultaneous, independent control of the scattering properties of different component pairs could be achieved by combining rf with one or more existing Feshbach resonances, as has been suggested theoretically~\cite{moerdijk96,zhang09, alyabyshev09} but not yet observed. Similar conclusions may be drawn from related experimental work involving optical frequencies~\cite{bauer09}.

In this Letter we report a significant step towards independent control of collisions between different component pairs, demonstrating several resonances in $^{87}$Rb that are tunable with both rf and magnetic field.
Our scattering theory analysis reproduces the experimental data in detail. We accurately measure and characterize each of the strongest underlying magnetically tunable Feshbach resonances, and find that the primary role of the rf in our system is to couple the bound states that give rise to these resonances. Since rf is easily manipulated, our technique could be used to switch scattering lengths rapidly and precisely without the need to change a magnetic bias field, avoiding the deleterious effects of eddy currents and finite servoloop bandwidths. This feature alone could find use in studies of nonequilibrium phenomena~\cite{barankov04}, varying mean field energies to tune qubit phases~\cite{Daley08}, spin squeezing~\cite{li08}, and matter-wave analogues of nonlinear optical systems~\cite{kevrekidis03}.

\begin{figure}[tb]
	\centering
	\includegraphics[width=0.95\columnwidth, clip]{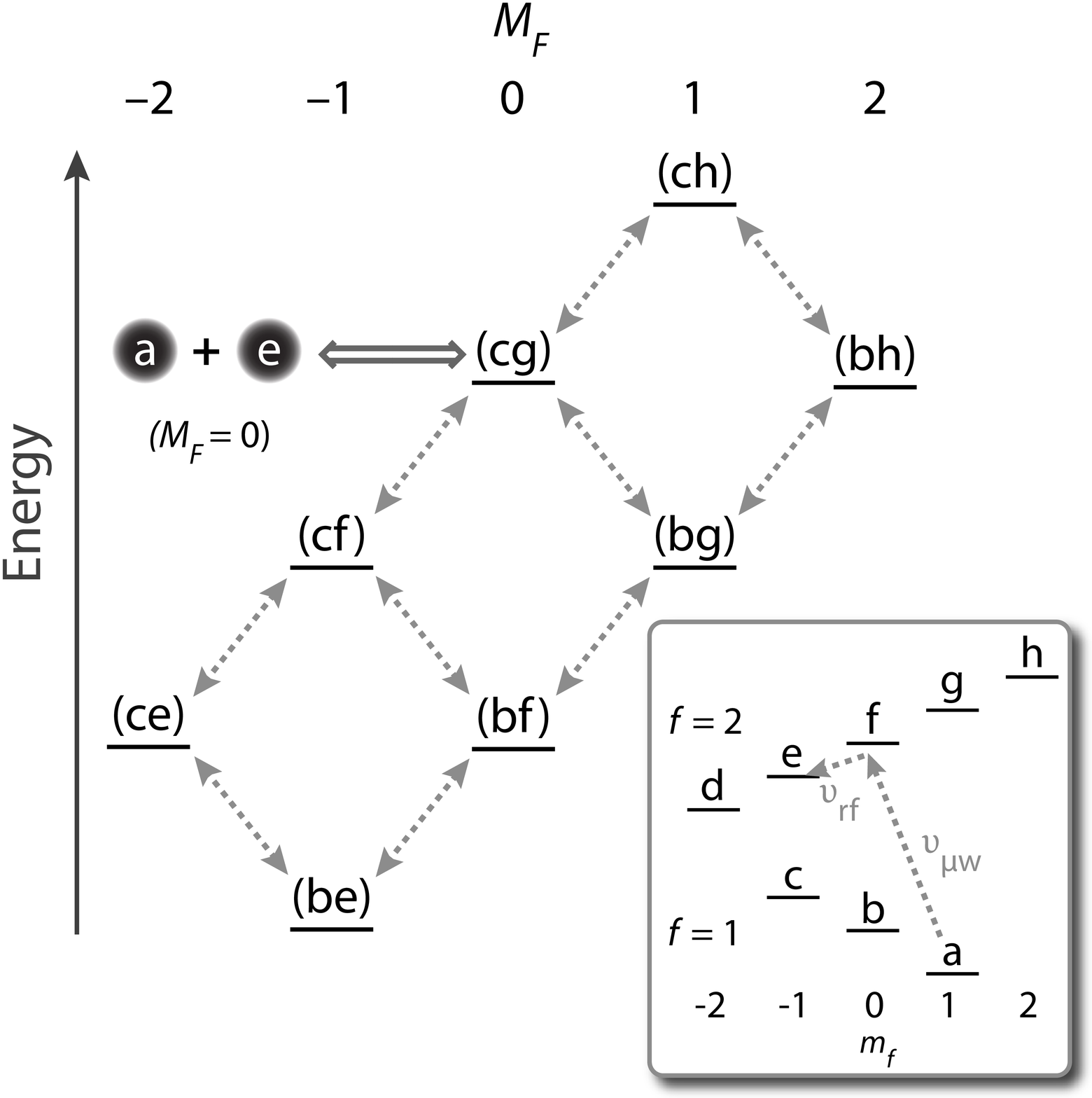}
	\caption{Schematic of rf-induced Feshbach resonances for the $M_F = 0$ $a + e$ entrance channel (black circles). At 9.1\,G. the $a+e$ channel is coupled by spin exchange to the $(cg)$ bound state (broad arrow). Seven other bound states are, in turn, coupled by rf induced magnetic dipole transitions (dashed arrows) to the $(cg)$ state. The rf photons, linearly polarized perpendicular to the magnetic field, can drive transitions with $\Delta M_F = \pm 1$.
	The inset shows the energies and labelings of the atomic Zeeman states of \Rb\ at a small, nonzero magnetic field, along with the two-photon $a \leftrightarrow e$ transition. The energy splitting between hyperfine levels $f =1,2 $ is not drawn to scale.
	The bound state labels are explained in the text.
	}
	\label{fig:coupling}
\end{figure}

Atomic collisions in the presence of a magnetic field may be described in terms of two-body channels, defined by the Zeeman state of each atom and the partial wave of their collision. We consider only $s$-wave collisions of $^{2}S$ \Rb\ atoms, and label the atomic Zeeman states alphabetically in order of increasing energy. The projection $m_f$ of the total atomic angular momentum $f$ is a good quantum number at any magnetic field $B$. At the magnetic fields of interest, $f$ is only approximately conserved. The link between the alphabetical and approximate $|f, m_f\rangle$ Zeeman state labels is sketched in the inset of \reffig{fig:coupling}.
In a collision between two atoms at nonzero magnetic field, $M_F = m_{f_1}+m_{f_2}$ is conserved~\cite{review}, where $M_F$ is the projection of the total angular momentum $\vec{F} = \vec{f}_1 + \vec{f}_2$.
We note that $F$ is only a good quantum number at zero field.
A Feshbach resonance arises when the threshold of the entrance channel, $\alpha + \beta$, where Greek letters indicate atomic Zeeman states,
is degenerate with a bound state of the same $M_F$~\cite{chin_review}.

The experimental apparatus is similar to that described in Ref.~\cite{mertes07}. We begin with a condensate of $2 \times 10^5$ \Rb\ atoms prepared in a magnetic trap in state $h$. We then transfer the condensate into a horizontal crossed-beam optical dipole trap consisting of mutually orthogonal 1064\,nm beams, each of waist $30\,\mu$m and power $22$\,mW. The potential experienced by the atoms is approximately cylindrically symmetric, and has measured radial and axial trap frequencies of $\{\omega_r,\omega_z\}=2\pi \times \{120,160\}$\,Hz. We apply a uniform magnetic bias field using three orthogonal pairs of Helmholtz coils. Microwave and rf signals (both locked to a frequency standard and with frequencies $\nu_{\mu \mathrm{w}} \sim 6.8$\,GHz and $\nu_\mathrm{rf}\sim$ a few\,MHz, respectively) are used to drive transitions between the atomic Zeeman states.
The magnetic field is initially calibrated by spectroscopy on the $a \leftrightarrow f$ and $a\leftrightarrow h$ atomic Zeeman transitions.

We first characterize the dominant Feshbach resonances in channels with $f_1 = 1$ and $f_2 = 2$ in the absence of rf radiation. Of these, the $a+e$ resonance near $B = 9.1$\,G (1\,G$\, = 10^{-4}$\,T) has been previously predicted~\cite{vankempen02} and observed~\cite{erhard04,widera04}. Each resonance location is found by preparing a superposition of the two Zeeman states in its entrance channel using combinations of rf and microwave pulses and sweeps. We then release the atoms from the optical trap and measure the number of atoms remaining in the $f=2$ hyperfine level after 15\,ms of expansion. Atoms are lost to inelastic spin relaxation to the energetically lower $(f_1 = 1) + (f_2 = 1)$ channels, as well as to three-body recombination, both of which peak near the location of the resonance~\cite{chin_review}.
To mitigate magnetic field variations (typically a few hundred $\mu$G over the course of a loss profile measurement) the data are taken in random order, and within each set we further interleave an accurate spectroscopic determination of the magnetic field through a measurement of the $a \leftrightarrow h$ transition. We are unable to discern any systematic differential mean-field or light shift in the $a+e$ resonance location when measuring loss profiles with far fewer (6--20\,$\times 10^3$) atoms, or in a single trapping beam, or with untrapped atoms.

The measured resonance locations and calculated resonance properties are summarized in Table~\ref{table:comparison}. The large number of resonances, and their grouping, arise because of the similarity of the scattering lengths of the singlet $X^{1}\Sigma_g^{+}$ and triplet $a^{3}\Sigma_u^{+}$ potentials of \Rb~\cite{review}. This causes each possible value of $F$ (1, 2, and 3) to have a zero-field least bound state with nearly the same energy. These bound states give rise to the 9\,G and 18\,G resonances. At these fields the Zeeman splittings are significantly larger than those between the zero-field states, so the states associated with the three values of $F$ mix strongly. Although bound states generally have components in several Zeeman channels with a common $M_F$, in this case we find that each has a dominant channel with admixture more than 90\%. We label these bound states $(\gamma \delta)$ for the corresponding channel $\gamma + \delta$.

\begin{table}[t]
\centering
\begin{tabular}{| c | c | c | c | c | c |}
\hline
\T
Resonance  & $B_0$ (G) & $B_0$ (G) & $a_\p{bg} / a_0$ &  $\Delta B$   & $\gamma_B $ \\
character & (exp.)  & (theory)    &                  &     (mG)       &   (mG)                 \\
[1ex]
\hline
\hline
\T
$a + d \leftrightarrow (cf) $ & 9.0918(5) & 9.093 & 98.0 & 1.3 & 4.6 \\
$a + e \leftrightarrow (cg) $ & 9.1047(5) & 9.105 & 97.7 & 2.0 & 4.7 \\
$a + f \leftrightarrow (ch) $ & 9.0448(5) & 9.045 & 97.7 & 1.3 & 3.0 \\
[1ex]
\hline
\T
$a + d \leftrightarrow (be)$ & 17.9208(2) & 17.914 & 98.0 & 0.95 & 3.4 \\
$a + e \leftrightarrow (bf)$ & 17.821(1)  & 17.808 & 97.7 &      &     \\
$a + f \leftrightarrow (bg)$ & 17.9848(2) & 17.975 & 97.7 & 1.4  & 3.2 \\
$a + g \leftrightarrow (bh)$ & 18.4108(5) & 18.418 & 98.0 & 3.8 & 12.8 \\
$b + d \leftrightarrow (ce)$ & 18.0059(5) & 18.002 & 97.4 & 3.5 & 6.1  \\
$b + e \leftrightarrow (cf)$ & 18.1707(6) & 18.172 & 98.3 & 1.4 & 15.5 \\
[1ex]
\hline
\end{tabular}
\caption{Feshbach resonances of $^{87}$Rb near $B = 9$\,G and 18\,G. The first column gives the entrance channel $\alpha + \beta$ and the character of the bound state $(\gamma \delta)$ causing the resonance (see text). The second column gives the experimentally determined resonance locations, with one standard deviation combined systematic and statistical uncertainties in parentheses. The remaining columns give resonance parameters fitted to the scattering lengths calculated with the technique of Ref.~\cite{hanna09}.
The parameters are determined from the formula $a(B) = a_\p{bg}[1 - \Delta B / (B - B_0 - i \gamma_B/2)]$, where $a_\p{bg}$ is the background scattering length, $\Delta B$ is the resonance width, and $\gamma_B$ is the decay width expressed in magnetic field units, representing inelastic spin relaxation. Finally, $a_0 = 0.05292$\,nm. The $17.821\,$G $a + e$ resonance is extremely narrow and could not be reliably fit for width. For the resonances shown here, the method of Ref.~\cite{hanna09} is accurate only to $\sim$10$^{-4}$, explaining the minor disagreement with some of the measured locations.}
\label{table:comparison}
\end{table}

We now consider how the Feshbach resonances are modified in the presence of rf radiation. We study collisions in the $a+e$ entrance channel near 9.1\,G, starting with a condensate of atoms entirely in state $a$. State $a$ is linked to state $e$ via a two-photon microwave + rf transition~\cite{widera04}, detuned a few hundred kHz from the intermediate state~\cite{noteIntState}. We realize adiabatic passage~\cite{mewes97} from state $a$ to state $e$ by sweeping the microwave frequency through the two-photon resonance. The rf radiation has a fixed frequency and plays a second role in modifying the scattering properties. During the sweep, the two-photon coupling creates superpositions of atoms in states $a$ and $e$, populating the entrance channel. Inelastic collisions between dressed atoms result in losses that we assess by counting the remaining $e$ atoms at the end of the sweep.

\begin{figure}[tb]
	\centering
	\includegraphics[width=0.95\columnwidth, clip]{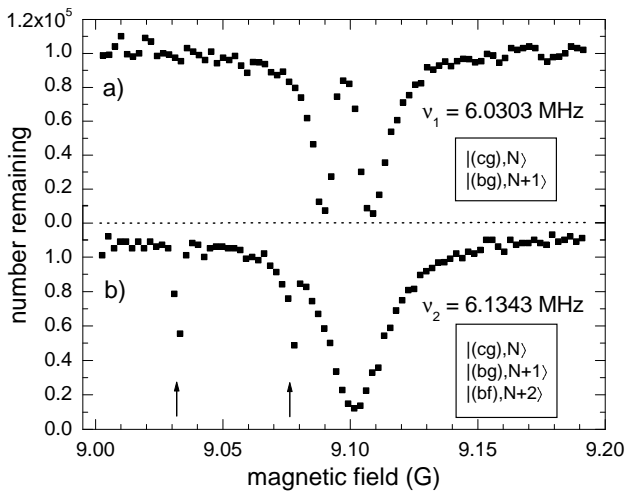}
	\caption{Atom loss data from the $a + e$ entrance channel in the vicinity of the 9.1\,G resonance with applied rf radiation. (a) For $\nu_1 = 6.0303$\,MHz the rf dressing splits the undressed 9.1\,G $a + e$ Feshbach resonance into two separated features as a result of an avoided crossing. (b) For $\nu_2 = 6.1343$\,MHz the coupling of three bound states leads to two narrow features (arrows) in addition to the primary loss feature. The bound states $\ket{(\gamma \delta),N'}$ that contribute to the loss features are listed in the boxes within each panel.}
\label{fig:avcross}
\end{figure}
\begin{figure}[bt]
	\centering
		\includegraphics[width=1.00\columnwidth]{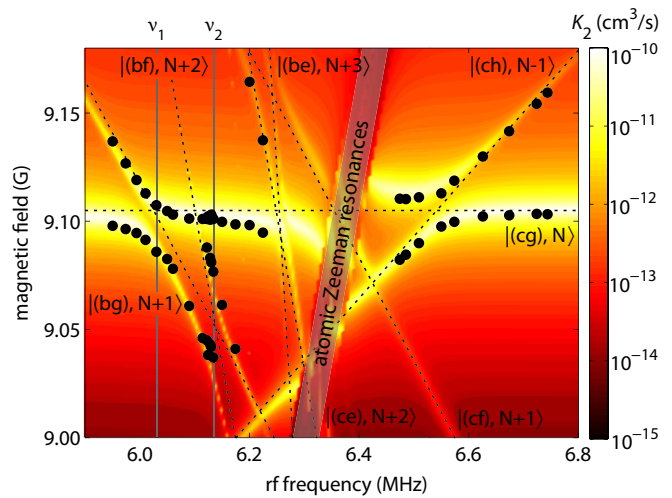}
	\caption{(color) Locations of experimental resonance features (solid points), and theoretical spin relaxation loss rate coefficient $K_2$ (pseudocolor intensity), for collisions in the $\ket{a+e,N}$ entrance channel as a function of magnetic field and rf frequency. Dotted lines indicate where the bound state assignments $\ket{(\gamma \delta), N'}$ cross the $\ket{a+e,N}$ threshold. Vertical lines labeled $\nu_{1,2}$ correspond to the loss profiles in \reffig{fig:avcross}. No data could be taken close to the region labeled ``atomic Zeeman resonances,'' where atomic transitions occur among several entrance channels. }
	\label{fig:K2_40}
\end{figure}

Two representative loss profiles indicating the role played by the rf radiation in modifying the scattering properties are shown in Fig.~\ref{fig:avcross}. We survey a 200~mG range about the location of the $a+e$ Feshbach resonance. For an applied rf frequency of 6.0303\,MHz, the loss feature due to the undressed Feshbach resonance is split, with losses strongly suppressed between the two features. For an rf frequency of 6.1343\,MHz we see three features, two of which are quite narrow. All of the features change location and character as the rf and magnetic field are varied, as shown in Fig.~\ref{fig:K2_40}.

To explain our data we have developed a model based on that presented in Ref.~\cite{hanna09}. An $a + e$ collision with $N$ photons in the rf field is linked by molecular spin-exchange interactions to a ``spin-exchange block'' consisting of all channels with the same $M_F$ and $N$. Radiofrequency dressing is then applied to each of these channels, now described by $\ket{\alpha+\beta, N}$. The radiation can drive several $(\Delta M_F = \pm 1, \Delta N = \pm 1)$ transitions from the entrance channel block containing $\ket{a+e, N}$. Spin exchange interactions occur for each channel coupled in by rf. Consequently, for each of these we include the entire spin-exchange block of which it is a part. For the present case, including channels up to three transitions from the entrance channel block is sufficient.
After calculating the rf-induced magnetic dipole coupling between these channels, we diagonalize the Hamiltonian at asymptotically large interatomic separation. This provides a basis of rf-coupled channels with which we can calculate all observable scattering properties by extending the simplified model of Feshbach resonances of Ref.~\cite{hanna09}.
The model exploits the fact that the collision energy of the atoms is much smaller than the characteristic energy scales of the interatomic potentials.
This allows us to express the short range scattering properties in an energy-independent way, and to use an approximate potential of the form $-C_6/r^6$ for all interatomic distances $r$, where $C_6$ is the van der Waals coefficient~\cite{gao05}. The entire calculation requires only known atomic parameters, the measured rf Rabi frequency, and three properties of the interactions: the singlet and triplet scattering lengths, and $C_6$~\cite{noteScatteringLengths}.

We compare our calculated spin relaxation loss rate coefficient $K_2$ to the experimental measurements of loss features in \reffig{fig:K2_40}. The locations of the loss features agree well with our experimental measurements. Minor disagreements occur for data close to the atomic Zeeman resonances, where the main $a+e$ loss peak appears at slightly lower fields than the theory predicts. These discrepancies may result from the approximate Feshbach resonance widths generated by our theory~\cite{hanna09}, the neglect of higher partial waves, or dressing due to the microwave photon.

The spin-exchange blocks included in our full calculation support the eight bound states shown in \reffig{fig:coupling}, which give rise to the resonances listed in Table~\ref{table:comparison}. The rf radiation couples these bound states together, since their energy spacings are approximately the same as those of the atomic transitions at 9.1\,G. This coupling leads to a sequence of avoided crossings between rf-dressed bound states. We observe pronounced losses as an rf-dressed bound state with a nonzero $(cg)$ component crosses the $\ket{a+e,N}$ threshold. It is important to note that, in general, the rf-induced bound-free and free-free couplings included in our calculation can be expected to be of significance to rf-dressed Feshbach resonances~\cite{noteli}. However, the many coincident bound states and the similarity of all background scattering lengths make these effects negligible in the present case. This allows our simple interpretation in terms of the coupled bound states of \reffig{fig:coupling}.

We explain the structure of loss features in \reffig{fig:K2_40} by identifying the corresponding bound states. Two of these avoided crossings are particularly pronounced. The first, at ($B = 9.1\,\text{G},\,\nu_{\text{rf}} = 6.02$\,MHz) arises from the crossing of the $\ket{(bg),N+1}$ and $\ket{(cg),N}$ states. This produces the Autler-Townes doublet
shown in \reffig{fig:avcross}a, in a manner analogous to that recently observed at optical frequencies~\cite{bauer09}. Similarly, the avoided crossing at $(B =9.11\,\text{G},\,\nu_{\text{rf}} = 6.55\,\text{\,MHz})$ arises from the crossing of the $\ket{(ch),N-1}$ and $\ket{(cg),N}$ states. The loss profile of \reffig{fig:avcross}b arises when three bound states are coupled at an rf frequency of 6.13\,MHz. Three rf-dressed bound states, each with a $(cg)$ bound state component, cross the $\ket{a + e,N}$ threshold at different magnetic fields, yielding three rf-dressed Feshbach resonances. The other observed loss features appear as further bound states of \reffig{fig:coupling} are coupled in by higher-order transitions.

Our calculations show that each rf-dressed Feshbach resonance produces not only a loss feature but also a tuning of the real part of the scattering length. Since the background scattering length of the $a + e$ channel is close to the critical value separating the regimes of miscibility in a binary quantum fluid, even small scattering length changes induced by rf radiation could result in radically different dynamical and ground-state properties~\cite{ho96}. We anticipate exploring this in future experiments.

In conclusion, we have coupled together several proximate Feshbach resonances using rf radiation. After  characterizing the strongest $(f_1 = 1) + (f_2 = 2)$ resonances near 9\,G and 18\,G, we observed a series of avoided crossings due to rf-dressed resonances. Our scattering theory analysis provided quantitative agreement with the experimentally measured locations of these resonances, and we developed an intuitive picture based on coupled bound states interacting with the entrance channel threshold.

P.S.J acknowledges partial support by the U.S. Office of Naval Research; the Amherst researchers acknowledge funding by the National Science Foundation. We thank K.~M. Mertes, D.~H. Guest, M.~L. Goldman, and D.~M. Bianchi for additional experimental work, and Stephen Maxwell for initiating our collaboration.

\bibliography{tomsrefs}

\begin{thebibliography}{31}
\bibitem[{not({\natexlab{a}})}]{noteIntState}
\bibinfo{note}{The rf Rabi frequency on the $ a \leftrightarrow b$ transition
  is $2\pi \times 38.6(5)$\,kHz, and the microwave Rabi frequency on the $ a
  \leftrightarrow f$ transition is typically $2\pi \times 10$\,kHz. We have
  used both $f$ and $b$ as intermediate states with identical results, although
  the former cannot be used for fields near the 9.0448\,G $a+f$ Feshbach
  resonance.}
\bibitem[{not({\natexlab{b}})}]{noteScatteringLengths}
\bibinfo{note}{The singlet and triplet scattering lengths were used as fit
  parameters in our model, which is optimized to reproduce the experimental
  resonance locations in Table~\protect\ref{table:comparison}.}
\bibitem[{not({\natexlab{c}})}]{noteli}
\bibinfo{note}{For example, we have calculated that rf bound-free coupling may
  be used to create resonances in $^{6}$Li. Details will be published
  elsewhere.}
\end{thebibliography}

\end{document}